\documentclass[12pt]{iopart}

\usepackage{graphicx}
\usepackage{iopams}
\usepackage{soul}
\usepackage{color}
\usepackage[section]{placeins}
\newcommand{\Wcm}{ W/cm$^2$ \hspace{1mm}}

\begin{document}
\bibliographystyle{unsrt}

\title[Towards the Optimisation of Direct Laser Acceleration]{Towards the Optimisation of Direct Laser Acceleration}
%The relative importance of channel fields in the existence of an optimal density for Direct Laser Acceleration 

% The influence of target density on channel fields for optimized DLA 

%\title[Optimizing Direct Laser Acceleration]{Direct Laser Acceleration of electron beams to \textcolor{red}{600~MeV} by optimizing plasma density}% Force line breaks with \\
%\thanks{Footnote to title of article.}

\author{A.~E.~Hussein$^{1,*}$,
A.~V.~Arefiev$^{2}$,
T.~Batson$^{1}$,
H.~Chen$^{3}$,
{R.~S.~Craxton$^{4}$},
A.~S.~Davies$^{4}$,
D.~H.~Froula$^{4}$, 
Z.~Gong$^{5}$,
D.~Haberberger$^{4}$, 
%O.~Jansen$^{2}$, 
Y.~Ma$^{1}$,
{P.~M.~Nilson$^{4}$,
W.~Theobald$^{4}$},
{T.~Wang$^{2}$}, 
{K.~Weichman$^{2}$}, 
{G.~J.~Williams$^{3}$},
{L.~Willingale$^{1}$}}

\address{$^1$Center for Ultrafast Optical Science, University of Michigan, Ann Arbor, MI 48109, USA}
\address{$^2$University of California San Diego, San Diego, CA 92093 USA}
\address{$^3$Lawrence Livermore National Laboratory, Livermore, CA 94550, USA}
\address{$^4$Laboratory for Laser Energetics, University of Rochester, Rochester, NY 14623, USA}
\address{$^5$Center for High Energy Density Science, The University of Texas, Austin, TX 78712, USA}
\address{$^*$Present address: Department of Electrical and Computer Engineering, University of Alberta, Edmonton, AB, T6G 2R3, Canada}

\ead{aehussein@ualberta.ca}

\begin{abstract}
Experimental measurements using the OMEGA EP laser facility demonstrated direct laser acceleration (DLA) of electron beams to (505 $\pm$ 75) MeV with (140 $\pm$ 30)~nC of charge from a low-density plasma target using a 400 J, picosecond duration pulse. Similar trends of electron energy with target density are also observed in self-consistent two-dimensional particle-in-cell simulations. The intensity of the laser pulse is sufficiently large that the electrons are rapidly expelled from along the laser pulse propagation axis to form a channel. The dominant acceleration mechanism is confirmed to be DLA and the effect of quasi-static channel fields on energetic electron dynamics is examined. A strong channel magnetic field, self-generated by the accelerated electrons, is found to play a comparable role to the transverse electric channel field in defining the boundary of electron motion.

% AEH April 9 
%, indicating the key role of the magnetic field in this regime, rather than the transverse electric field within the channel. 

%Experimental measurements using the OMEGA EP laser facility demonstrated direct laser acceleration (DLA) of electron beams to (250 $\pm$ 10) MeV and $\sim$ 170 nC of charge from a low-density plasma target using a high-energy, picosecond duration pulse.
%Fully self-consistent two-dimensional particle-in-cell simulations were in good agreement with experimental results, confirming DLA as the dominant acceleration mechanism and elucidating the effect of quasi-static channel fields on energetic electron dynamics.
%A strong channel magnetic field is found to define the boundary of electron motion, indicating the dominant role of the magnetic field in DLA, in contrast with previous work that has focused on channel electric fields. 

\end{abstract}

\maketitle

\section{Introduction}

Modern laser technology and the realization of high-intensity, short-pulse laser systems using chirped-pulse amplification \cite{strickland1985compression} has expanded the frontiers of physics for fundamental research and novel technological applications including laser-based schemes for charged particle acceleration. In all laser-plasma interactions, the pivotal step and basis of all subsequent phenomena is governed by the transfer of energy between the laser fields and plasma electrons. The generation of copious, high-energy electrons is key for driving secondary particle and radiation sources, such as energetic ions \cite{snavely2000intense,wilks2001energetic,willingale2006collimated}, hard X-rays \cite{chen2013mev,albert2016applications,stark2016enhanced} neutrons \cite{lancaster2004characterization,roth2013bright} and electron-positron beams \cite{chen2009PRL,Vranic2018SciRep}.  Elucidating and optimizing the dynamics of electron heating and acceleration for different regimes of plasma density and laser pulse duration is central to the development of these sources.

A laser field can propagate through a plasma if the electron density $n_e$ is below the critical density, $n_{crit} \equiv m_e \omega_0^2/(4 \pi e^2)$, where $m_e$ is the electron mass, $\omega_0$ is the laser frequency and $e$ is the electron charge. In this regime, the mechanism for laser-driven electron acceleration is highly dependent on laser pulse duration ($\tau_{L}$).
The dominant electron acceleration mechanism can be inferred from the relationship between $\tau_{L}$ and the plasma frequency, $\omega_{pe} = \sqrt{4 \pi n_e e^2/m_e}$. When $\tau_{L} \simeq 1/\omega_{pe}$, as is typical for femtosecond-duration pulses and low-density targets, laser wakefield acceleration (LWFA) dominates, and electrons can be accelerated up to many-GeV energies by the longitudinal electric field of electron plasma waves \cite{Tajima1979}. The wakefield structure forms because electrons within the focal region of the laser pulse experience the ponderomotive force, expelling them from regions of high intensity to form a cavity containing the heavier ions. Once the laser pulse passes, the electrons return towards the axis, and the wakefield structure is formed. LWFA electron beams can be high-energy (many GeV) \cite{Downer,kim2013enhancement,leemans2014multi,gonsalves2019petawatt}, mono-energetic \cite{Mangles_Nature_2004,Geddes_Nature_2004,Faure_Nature_2004} and low divergence (on the order of a few milliradians \cite{albert2013angular}), however, the total charge of the electron beam is typically low, on the order of tens of picocoulombs \cite{leemans2016limits}. Higher charge electron beams are preferable for the generation of secondary sources. 

For picosecond (ps) duration pulses, the laser pulse duration is typically much greater than the plasma period. At low intensities, ps pulses can accelerate electrons \textit{via} self-modulated laser wakefield acceleration (SM-LWFA), which has been shown to produce electron beams with charge on the order of tens of nanocoulombs (nC) \cite{modena1995electron,najmudin2003self,lemos2018bremsstrahlung}.
However, as the laser intensity is increased (typically above 10$^{18}$ \Wcm for a $\lambda_0 = 1$ $\mu$m laser), the sustained ponderomotive force means the electrons are unable to return into the electron depleted region, therefore a wakefield is unable to form (except perhaps at the rising intensity of the leading edge of the laser pulse), and instead an ion channel is established. Within this channel, strong radial space charge fields can be present and direct laser acceleration (DLA) mechanisms become dominant. Eventually, the radial electric field leads to a ``Coulomb explosion'' of the ions \cite{krushelnick1999multi,sarkisov1999self}, reducing the strength of the radial electric fields of the ion channel. 

The basis of DLA is the transfer of energy directly from the laser to plasma electrons, with electrons gaining longitudinal momentum through the $\mathbf{v} \times \mathbf{B}$ force \cite{pukhov1998relativistic,pukhov1999particle}. In vacuum, the maximum energy an electron initially at rest can gain directly from a plane electromagnetic wave, with normalized amplitude increasing from zero to $a_0$ is given by: $\gamma_{vac} = \left( 1 + a_0^2/2 \right)$ \cite{arefiev2016beyond}, where $a_0 = |e|E_0/\left( m_e c \omega \right)$ is the normalized amplitude of a laser pulse with electric field $E_0$ and $\gamma$ is the relativistic factor. In this vacuum case, a $\lambda_0 = 1$ $\mu$m wavelength pulse with $a_0$ = 7 should yield $\gamma_{max} = 25.5$ ($\epsilon_{max} \simeq 13$ MeV). However, experimental measurements of DLA in a plasma have demonstrated electron energies vastly exceeding this limit \cite{gahn1999multi,Mangles2005,willingale2013surface}. 

Enhanced electron energy is largely attributed to the formation of the ion channel through the ponderomotive expulsion of electrons in the transverse direction. This channel evolves on the ion timescale and is associated with transverse and longitudinal electric fields that are quasi-static relative to the timescale of electron motion. While these fields are much weaker than the laser field, they can have a profound impact on the dynamics of electrons injected into the channel \cite{shvets1994instabilities,pukhov1999particle,robinson2013generating,arefiev2015novel,arefiev2016beyond,khudik2018far}, in particular by mitigating electron dephasing from the laser pulse. Under specific conditions in which the electron oscillation frequency matches the laser frequency, a resonance effect has been postulated to occur, increasing the transverse momentum of the electron, which is then transformed into longitudinal momentum through the $v \times B$ force. \cite{pukhov1999particle}. However, it has been previously shown \cite{arefiev2012parametric} that the dynamics of an electron irradiated by a plane wave within a static ion channel are non-linear, with strongly modulated eigenfrequency leading to a threshold process rather than a linear resonance. Strong quasi-static azimuthal magnetic fields are also generated through the driving of longitudinal electron currents by the intense laser pulse \cite{stark2016enhanced,jansen2018leveraging,wang2019structured}. A sufficiently strong azimuthal magnetic field may play a role in reinjecting an escaping electron into the beam volume to undergo further acceleration \cite{pukhov1998relativistic}. The impact of these fields has been largely neglected in favor of quasi-static electric fields in the context of DLA. 

While DLA electron beams are typically broadband and of lower peak energy than LWFA beams \cite{gahn1999multi,Mangles2005}, this mechanism can produce high-charge electron beams ($\sim$~100s of nC \cite{santala2001observation,Mangles2005,willingale2013surface,ma2018ultrahigh}). These high-charge beams can have important applications for secondary radiation sources \cite{Kneip2008PRL}. Currently, many High Energy Density Science (HEDS) facilities, such as the National Ignition Facility at the Lawrence Livermore National Laboratory, the OMEGA Laser System at the University of Rochester Laboratory for Laser Energetics, and Laser Mégajoule at the Commissariat à l'Énergie Atomique, are coupled to a kilojoule-class short-pulse beam with ps duration. The optimization of electron acceleration and X-ray generation for radiographic probing of HEDS experiments using these pulses motivates further studies of electron acceleration mechanisms in this regime.

Additionally, an experimentally validated model of density optimisation for DLA has not yet been presented. The complexity in parameterizing DLA lies in the dynamic interplay of the oscillating laser field with quasi-static channel electric and magnetic fields. Early theoretical work has suggested scalings of DLA efficiency with laser intensity, channel length and interaction time \cite{pukhov1999particle,meyer1999electron,khudik2016universal}. These previous studies have largely focused on the mitigation of electron dephasing to increase energy gain \cite{pukhov1999particle,arefiev2012parametric,robinson2013generating,arefiev2015novel,khudik2018far}. However, the present models of electron acceleration under the action of laser and channel fields are highly dependent on variations in the electron initial momentum and phase, electron injection, and propagation instabilities \cite{arefiev2016beyond}. 

In this work, comparison of experimental results with fully self-consistent 2D particle-in-cell (PIC) simulations highlights the contribution of the dynamic, quasi-static channel magnetic field as well as the transverse electric channel field on electron acceleration. We present experimental measurements of electrons accelerated by the OMEGA Extended Performance (EP) laser system through the interaction of a 1.0 ps duration laser pulse with an underdense hydrocarbon (CH) plasma plume. An optimal plasma density for electron acceleration by DLA, producing electron beams with energies up to (505 $\pm$ 75) MeV and up to (140 $\pm$ 30) nanocoulomb charge, was experimentally observed. 2D PIC simulations demonstrate similar trends of electron energy gain as a function of target density, providing insight into key phenomena governing electron acceleration in this regime. Further, 2D PIC simulations demonstrate that the magnetic field of the plasma channel plays an important role in the confinement and subsequent acceleration of plasma electrons by the laser field.

\begin{figure}
\centering
\includegraphics[width=1\columnwidth]{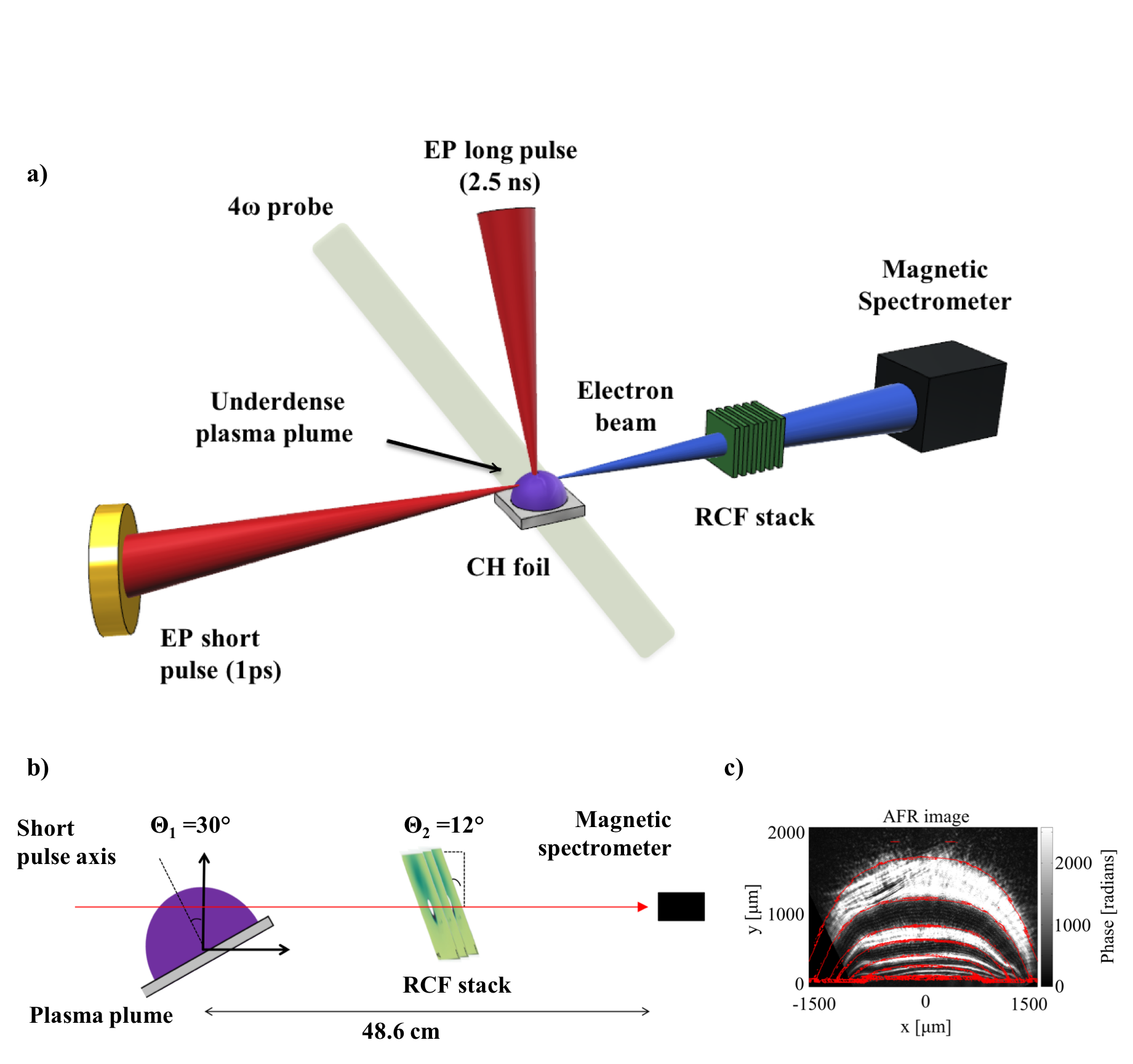}
\caption{\textbf{a)} Schematic of the experimental configuration showing the generation of an underdense plasma plume using a long-pulse heater beam, a 1.0 ps beam for electron acceleration, a 263 nm optical probe beam, and the location of beam diagnostics. \textbf{b)} Layout of radiochromic film and the magnetic spectrometer for diagnosing the accelerated electron beam (not to scale). \textbf{c)} An angular filter refractometry (AFR) image used to extract the plasma density profile.}
\label{setup}
\end{figure}

\section{Experimental setup}

Experiments were performed at the OMEGA EP laser system at the University of Rochester Laboratory for Laser Energetics. 
A schematic of the setup is given in Fig.~\ref{setup}a).
An underdense plasma target was produced using a single long-pulse UV heater beam (2.5 $\pm$ 0.3 ~ns pulse duration, 1214.6~$\pm$ 15.6 J of energy, $\lambda_0$~=~351 nm) with an 800~$\mu$m diameter super-Gaussian spatial profile focal spot incident on a flat CH foil ($125~\mu \rm{m}$ thickness).
Electron acceleration was driven using a (1.0 $\pm$ 0.1) ~ps full-width-half-maximum (FWHM) duration laser pulse, with a central wavelength of $\lambda_0$~=~1.053~$\mu$m and an average pulse energy of (414.6~$\pm$ 6.9)~J.
An $f/2$ off-axis parabolic mirror focused the light onto the edge of the plasma target, yielding a peak normalized vector potential in vacuum of $a_0 \simeq 7.0$.
The electron beam pointing and divergence were recorded using a stack of radiochromic film (RCF), positioned along the axis of the ps laser pulse at a distance of 8 cm behind the focal plane, as shown in Fig.~\ref{setup}b). The approximately 16 mm thick stack consisted of 10 sheets of HD-V2 film followed by two sheets of MD-V2-55 film interleaved with aluminium filters, with a 100 $\mu$m aluminium filter at the front. The RCF stack was tilted $12^{\circ}$ from normal to prevent back reflection of the laser. A hole in the center of the RCF stack allowed a direct line of sight to an absolutely calibrated magnetic spectrometer (EPPS \cite{Chen_RSI_2008}) 48.6 cm away from the focal plane for measurements of the electron spectrum along the axis of the main interaction beam.

The plasma density was varied by changing the interaction height of the 1.0~ps laser pulse above the plane of the CH foil, within a range of (1.5 -- 2.0)~mm. The timing between the ns and ps beams was 1.7~ns for the lowest density presented here and 2.5~ns for all others. The plasma density was measured by angular filter refractometry (AFR) \cite{haberberger2014measurements}, with example data shown in Fig.~\ref{setup}c). A fit to the data was found such that the plasma density profile can be approximated as a Gaussian function in two dimensions. In these experiments, peak plasma densities, $n_0$, ranging between (0.0095 - 0.11)$~n_{crit}$ were investigated, where $n_{crit}$ = 1.0 $\times$ 10$^{21}$ cm$^{-3}$ for $\lambda_0$ = 1.053 $\mu$m. The quoted density values refer to the peak density along the axis of the short-pulse laser in Fig.~\ref{setup}b). Given the interaction with the plasma plume at least 1~mm from the target and at late times in its evolution ($>$ 1.5 ns), we anticipate reasonable reproduction in density gradients and estimate the maximum variation in plasma length to be on the order of 0.5 mm. The experimentally determined plasma density profiles were compared with predictions from the two-dimensional hydrodynamic code SAGE [40]. In the region of interest, SAGE simulations differed from AFR measured densities by 30\% (overestimated in SAGE) 2.5~ns in the plasma evolution.

\section{Particle-in-cell simulations}

\begin{figure}
\centering
\includegraphics[width=1\columnwidth]{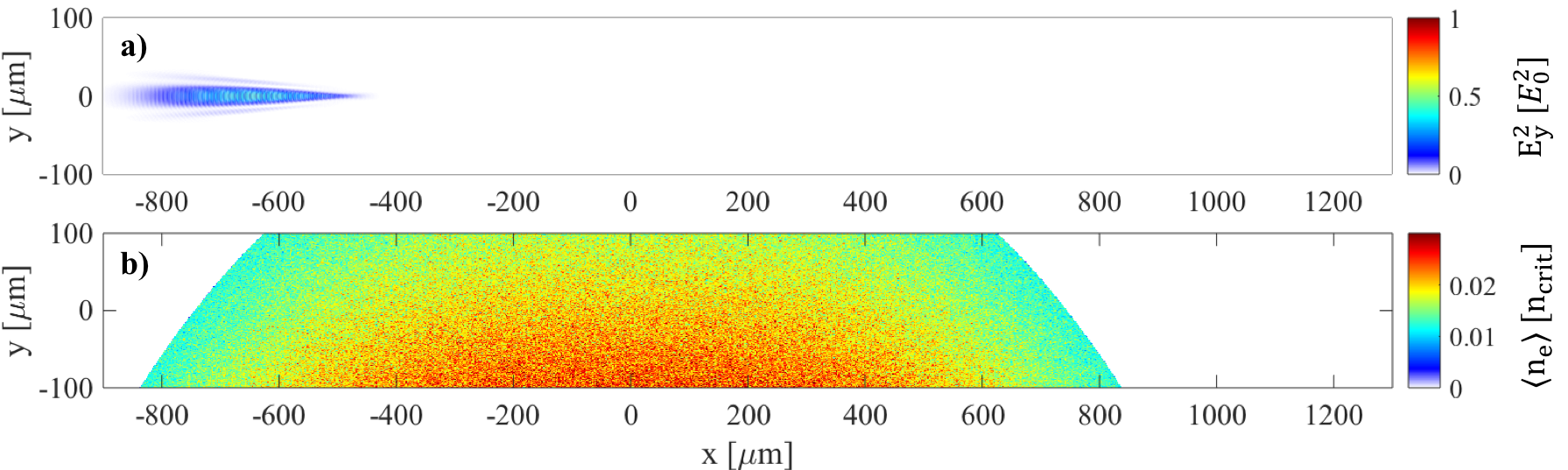}
\caption{\textbf{a)} Snapshot of the normalized laser intensity in vacuum, at $t=-0.5~$ps, propagating from left to right. \textbf{b)} Simulated plasma density profile, informed by AFR measurements, for a plasma characterized by $n_0=0.02~n_{crit}$.}
\label{sim_setup}
\end{figure}

Two-dimensional PIC simulations using the \textit{EPOCH} code \cite{arber2014epoch} (version 4.17.9) were performed to examine a laser pulse at relativistic intensity interacting with a plasma of sub-critical density. The simulations were designed to match the conditions of the OMEGA EP laser system. The 1.053~$\mu$m wavelength pulse was linearly polarized in $y$, and propagated in $x$. The time profile of the laser intensity was $\sin^2(\pi t/\tau)$ with a $\tau_L = 1.0$~ps FWHM duration ($\tau=2.0~\tau_L$). Two co-incident pulses and focal spots were used to approximate the experimental energy distribution in the focal plane: spot sizes of 3.4~$\mu$m and 17~$\mu$m, with laser intensities $I$ = 3.78 $\times$ 10$^{19}$ \Wcm and $I$ = 2.81 $\times$ 10$^{18}$ W/cm$^2$, respectively, corresponding to vacuum normalized vector potentials $a_0$ of 5.5 and 1.5. 

The simulation box was (2200 $\times$ 200) $\mu$m, spanning $x$ = [-900,1300] $\mu$m and $y$ = [-100,100] $\mu$m, with 30 cells per $\lambda$ in $x$, and 6 cells per $\lambda$ in $y$ and three macroparticles per cell for both electrons and ions. As shown in Fig.~\ref{sim_setup}, the laser entered the box at $y$ = 0, propagating from left to right, and traveled through vacuum before coming to focus in the plasma at $x= 490~\mu$m. The peak plasma density, $n_0$, was scaled from the profile extracted from AFR measurements (Fig.~\ref{setup}c)) to yield peak densities of (0.005 - 0.1)$~n_{crit}$ along the laser trajectory. Here we assume the same plasma length at each density. In a vacuum simulation, the laser reached peak intensity at a distance of 410~$\mu$m into the simulation box and a time referenced as $t = 0$~ps. Fully ionized carbon ions were treated as mobile and open boundary conditions were employed. Simulations were run at least until the accelerated electron beam exited the simulation box. Up to 10~ps of interaction time was simulated. A vertical probe plane placed at $x$ = 1295 $\mu$m in the simulation box recorded the positions, momenta and weight of all electrons with energy exceeding 10 MeV passing through the plane in the laser propagation direction (i.e. moving right). The electron and carbon densities, electromagnetic fields, current and particle locations were recorded every 250 fs, and time-averaged over five laser periods. Subsequently, particle tracking was conducted for time intervals from (-0.25 to 6.25) ps, with outputs of fields, density, electron position and momentum every 25 fs. 

The aim of these 2D simulations is to investigate experimental trends and to illustrate the physics of electron acceleration using picosecond duration laser pulses in underdense plasma, rather than for direct comparison with experimental results. While effects like diffraction and self-focusing may be underestimated in 2D simulations, recent work has shown that 2D simulations in this regime are qualitatively similar to 3D simulations \cite{gong2020forward}, and therefore reasonably capture the key physical phenomena relevant for interpretation of our experiments.

\section{Results and Analysis}

\subsection{Electron acceleration}

\begin{figure}
\centering
\includegraphics[width=1\columnwidth]{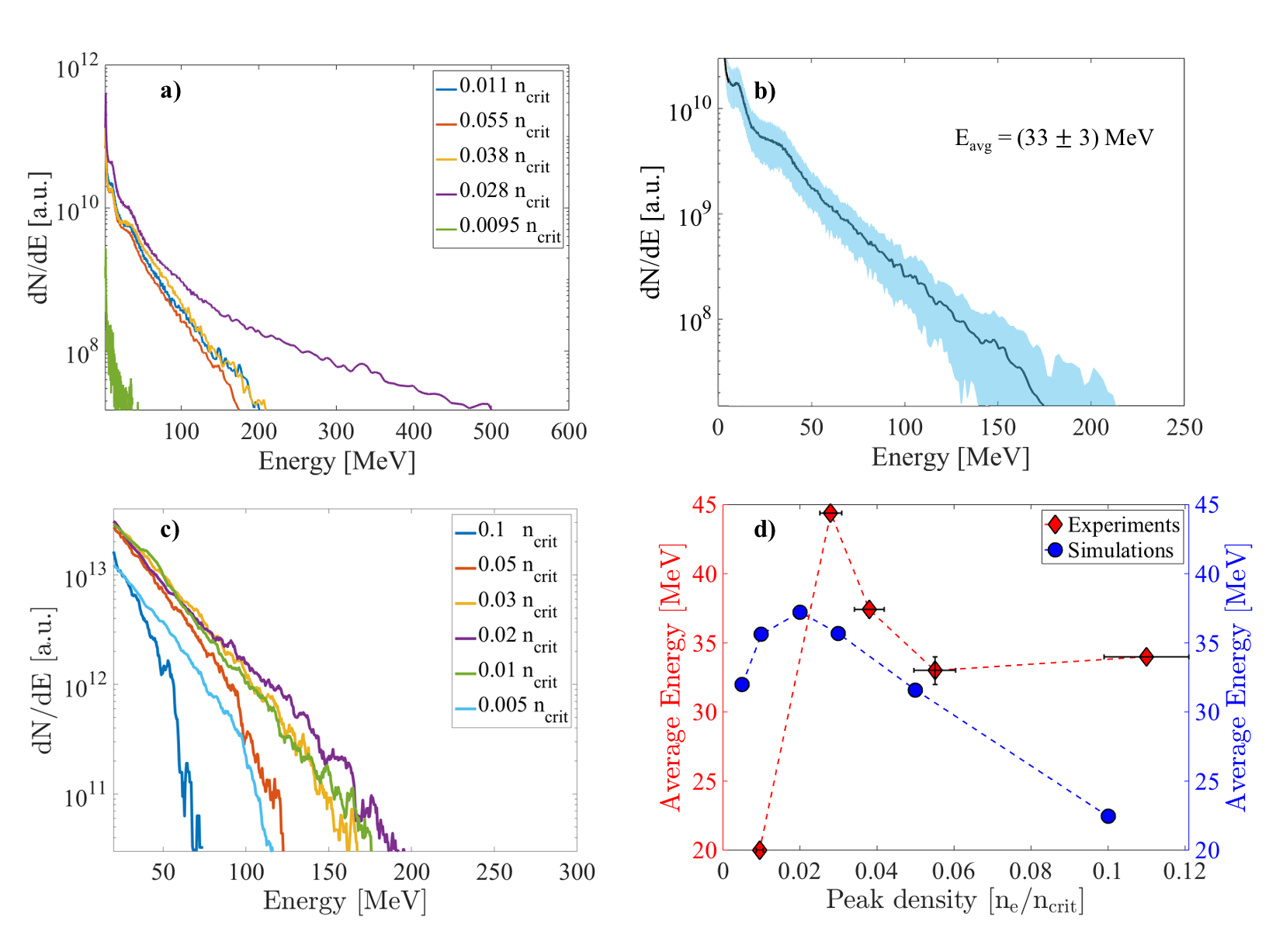}
\caption{\textbf{a)} Experimental spectra of escaped electrons for different peak plasma densities. \textbf{b)} Shot-to-shot variation in measured electron spectrum over five shots obtained at an estimated peak plasma density of $n_0=0.055~n_{crit}$. \textbf{c)} Total electron spectrum collected outside of the plasma at a probe 5~$\mu$m from the end of the simulation box, representing the beam exiting the plasma. \textbf{d)} Comparison between experimental and simulated average electron energies at the probe, showing good qualitative agreement (the left axis corresponds to experimental data).} 
\label{spectra}
\end{figure}

Experimental electron energy spectra from five different plasma densities are shown in Fig.~\ref{spectra}a). Significant acceleration of electron beams with a Maxwellian distribution extending to ($505 \pm 75)$~MeV is observed at a plasma density of 0.028$~n_{crit}$, indicating the existence of an optimal density for the generation of energetic electron beams. The electron spectra are shown to be reproducible at nominally identical experimental conditions ($n_0$ = 0.055$~n_{crit}$) with the average over five shots plotted in Fig.~\ref{spectra}b), where the shaded region represents the standard deviation.
In 2D PIC simulations, the escaping electron beam for electron energies $>$ 10 MeV was diagnosed outside of the plasma, as the electrons passed through the probe at $x$ = 1295 $\mu$m (Fig.~\ref{spectra}c), for comparison with the experimentally measured beam. While the existence of an optimal density in the simulations is not as dramatic as that observed in experiments, an optimal density for electron acceleration is also observed in simulations, with the highest energy beams produced at $0.02~n_{crit}$.

The average electron energy, evaluated from (10 - 300) MeV, is plotted for both simulations and experiments in Fig.~\ref{spectra}d). According to Ref. \cite{haberberger2014measurements}, the total error in the plasma density calculation using AFR is about $\pm$15\%. Experiments produced electron beams with a maximum average energy of (44 $\pm$ 3) MeV at 0.028$~n_{crit}$. For 0.055$~n_{crit}$, the electron spectrum is averaged over data from the five repeated shots of Fig.~\ref{spectra}b), yielding an average energy of ($33 \pm 3$)~MeV, where the quoted error reflects the standard deviation from five repeated experiments. The average electron energy appears to plateau for the highest densities in the experiments, however this trend is not reproduced in simulations, potentially owing to the underestimation of self-focusing and filamentation effects in 2D.

\subsection{Beam divergence, pointing and total charge}

\begin{figure}
\centering
  \includegraphics[width=1\columnwidth]{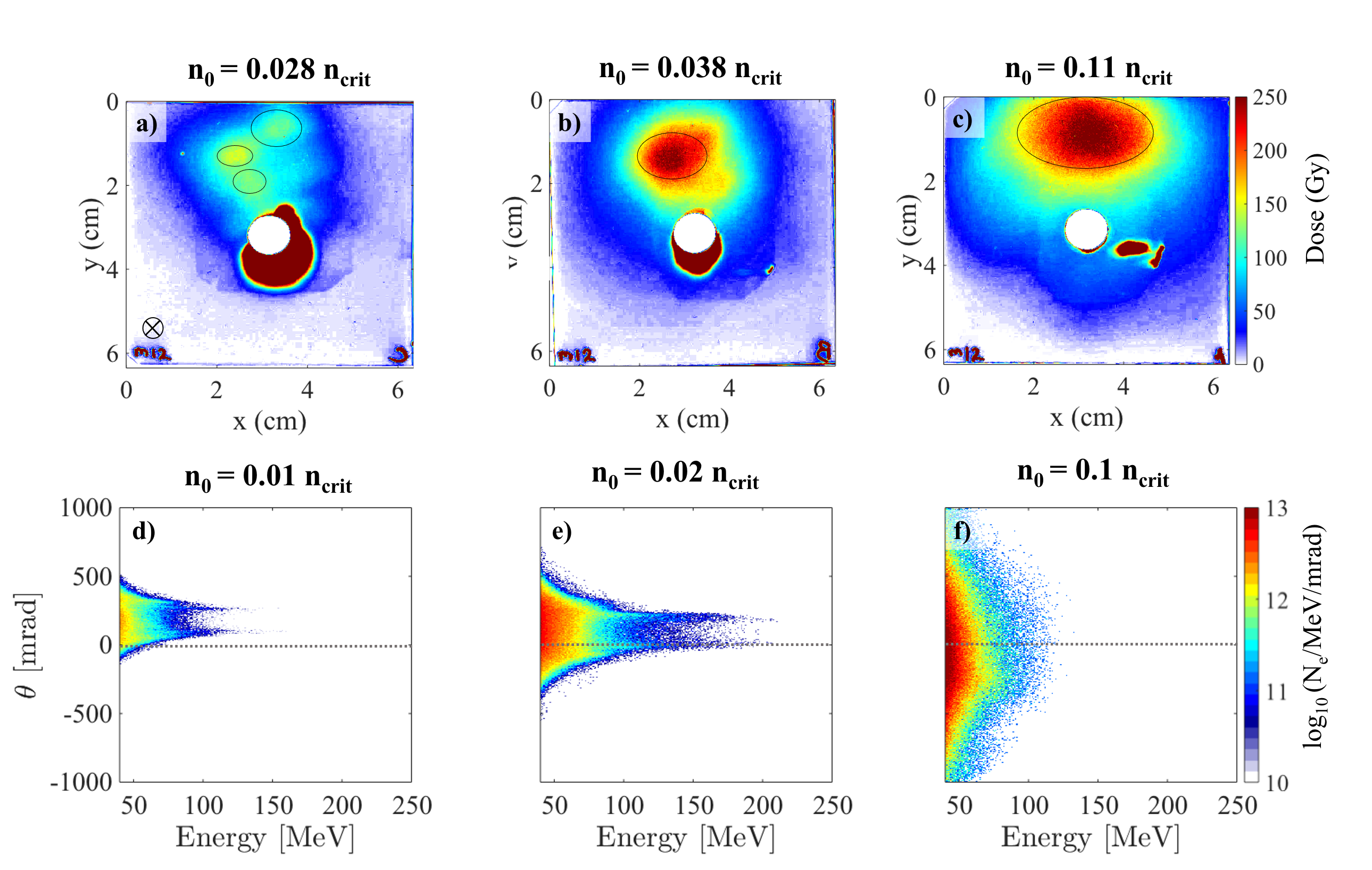}
\caption{\textbf{a-c)} Radiochromic film (RCF) images, along the axis of the laser beam for three plasma densities, shown in deposited dose, serving as a diagnostic of beam pointing and divergence. The hole in the center of the RCF stack is aligned with the 1.0~ps main interaction beam. The signal around the hole is due to line-of-sight radiation and is therefore ignored in calculations of total charge. \textbf{d-f)} Electron angular energy distribution from 2D simulations at $t$ = 1.75 ps for electrons with energy $>$ 20 MeV.}
\label{divergence}
\end{figure}

A stack of RCF positioned along the laser axis of the ps pulse provided information about the pointing, divergence and charge of the resultant electron beam (see Fig.~\ref{setup}b).
Scans of the final layer of MD-v2-55 film at the rear of the stack are shown from three different densities in Figs.~\ref{divergence}(a-c), in which the raw RCF signal was converted to dose, following Ref.~\cite{chen2016absolute}). The assumed center of the electron beam is indicated by ellipses in Figs.~\ref{divergence}(a-c). {Similar behavior has previously been attributed to space-charge-induced ion motion that can seed hosing-type instabilities \cite{Nilson2010NJP}. Here, no such hosing is observed at low density in simulations. At low density, the formation of beamlets is reminiscent of forking in the electron beam at high energies, which has previously demonstrated as a characteristic of DLA \cite{shaw2018experimental,gong2020forward}. Such forking is also observed in our simulations, presented in Fig.~\ref{divergence}(d-f), where the angular energy distribution is plotted for electrons with energy greater than 20 MeV.}

In all cases, the centroid of the electron beam or beamlets in Fig.~\ref{divergence}(a-c) lies above the original laser axis (centered approximately on the RCF hole) by about (1.75 - 2.5)~$\pm$~0.25~cm, or (212 - 297)~$\pm$~30~mrad for the lowest to highest density. These results indicate that the highest energy electrons may not be directed towards, or measured by, the magnetic spectrometer, which has a line-of-sight through the hole in the RCF film. The perturbation of the electron beam from the laser axis may be due to refraction of the laser pulse in the plasma gradient of the plume towards regions of lower density. 

The upward refraction of the laser beam is also present in simulations, evidenced by the angular distribution of the most energetic electrons in Figs.~\ref{divergence}(d-e) above the laser axis ($\theta=0$). At the highest density (Fig.~\ref{divergence}f), the angular distribution of energetic electrons appears to be nearly centered on the axis of the laser pulse; however, propagation instabilities such as filamentation are most severe at high density, and can significantly impact beam pointing.  Simulations indicate that deflection of the electron beam from the laser axis may also be due to the formation of sheath fields  \cite{willingale2006collimated,peebles2018high} as the beam exits the plasma.

The beam divergence as a function of plasma density from experiments was calculated by applying a Gaussian fits to the electron beam profiles on the RCF in Figs. \ref{divergence}(a-c), ranging from about (300-400) mrad FWHM with increasing with plasma density. Similar trends were reproduced in simulations. Multiple beamlets at low density were considered as a single beam for comparison. The total charge within the electron beam was estimated by determining the total number of electrons with energy $>$ 2 MeV reaching the electron spectrometer and assuming uniform distribution over the full solid angle of the beam profile, defined by its FWHM. Spurious signals near the hole, due to straight-through radiation and visible in Figs.~\ref{divergence}(a-c), were ignored. At the lowest density (0.0095$~n_{crit}$), there was insufficient signal on the RCF above background to make estimates, so no estimates for this density are provided. The number of electrons measured on the electron spectrometer for this density was three orders of magnitude lower than observed at 0.028$~n_{crit}$. 

The highest estimated charge beam, reaching (140 $\pm$ 30) nC, was obtained at the optimal plasma density (0.028$~n_{crit}$). The total charge in the beam for [0.11, 0.055, 0.038]$~n_{crit}$ were [111, 64, 70]~nC, respectively, with a standard error of $\pm$30~nC defined by the variation from five repeated shots at 0.055$~n_{crit}$. {These charge estimates are considered as an upper bound on the total charge in the electron beam, as they do not take into account spatial variation along the beam profile.} However, the measured beam charge is lower than previous results \cite{willingale2013surface}, which may be due to the presence of the RCF stack along the axis of the accelerated electron beam. Using the beam charge estimates and average energy in the electron beam, the conversion efficiency into electrons with energy greater than 10 MeV was estimated to reach a maximum of (0.48 $\pm$ 0.2)\% at 0.028$~n_{crit}$. Given the variation in electron beam pointing observed in Figs.~\ref{divergence}(a-c), the highest energy electrons may not be measured on the electron spectrometer, thereby reducing the average electron energy and resulting in an underestimation of the conversion efficiency.

\subsection{Channel formation}

\begin{figure}
\centering
    	\includegraphics[width=1\columnwidth]{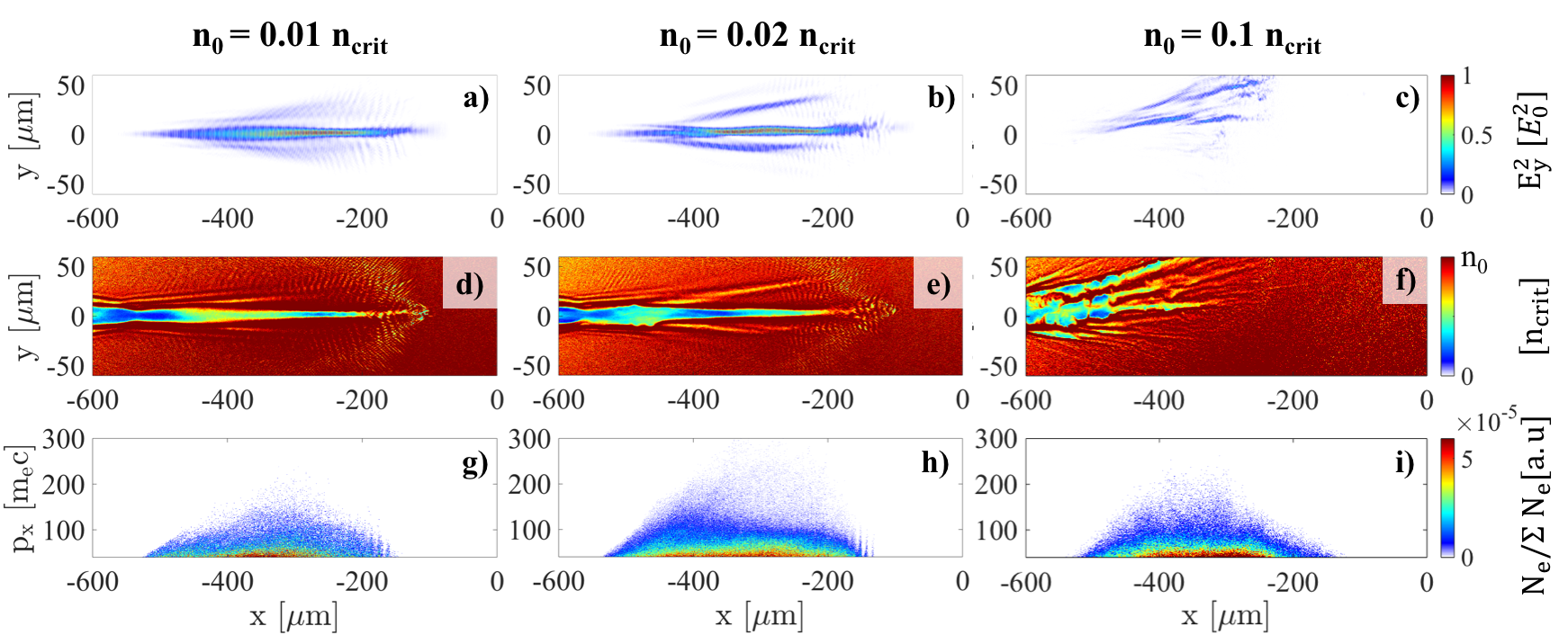}
\caption{Series of 2D PIC snapshots in the $(x,y$)-plane at $t$ = 0.75 ps. The laser enters the box at height $y$ = 0 and propagates from left to right. \textbf{a-c)} The laser intensity, with $E_y$ normalized to $E_0$ = 5.63 $\times 10^{8}$ statV/cm. \textbf{d-f)} The electron density, averaged over 5 laser periods. \textbf{g-i)} The electron phase space density $N_e$ for electrons with energy exceeding 20 MeV, where $p_x$ is the longitudinal momentum in arbitrary units. Note that these snapshots have been cropped from the full simulation window.}
\label{channels}
\end{figure}

Two-dimensional simulations provide insight into laser propagation effects and the role of the quasi-static plasma channel on electron acceleration at different plasma densities. Snapshots of the laser intensity, electron density, and phase space density are given in Fig.~\ref{channels} at a simulation time of 0.75 ps, as the laser self-focuses in the plasma plume. It is clear that the plasma density plays an important role in laser self-focusing and instability growth. At the lowest plasma densities, $0.01~n_{crit}$, a clear channel is formed, but is associated with moderate electron acceleration (c.f. Fig.~\ref{spectra}c). Additionally, at low densities the laser can expel all of the electrons from within the plasma channel, after which it is impossible for electrons to be injected and accelerated due to the comparatively high strength of the ponderomotive force \cite{robinson2013generating}. Channel formation is also evident at $0.02~n_{crit}$, with moderate filamentation of the laser pulse occurring with enhanced self-focusing relative to the lower density, but not impacting the ultimate formation of a channel propagation through the plasma at later times. At the highest density, $0.1~n_{crit}$, electrons are stochastically accelerated in the first few picoseconds of the interaction (c.f. Fig. \ref{tracking}c)). Subsequently, the propagation becomes unstable, resulting in filamentation and transverse break-up of the laser pulse. When a plasma channel cannot be formed, due to high levels of filamentation as observed at high density, there is no guiding of electron beams for enhanced electron energy gain from the laser field. However, instability growth at the beginning of the interaction can stochastically accelerate electrons, potentially impacting electron injection and pre-acceleration.

The phase space density of electrons with energy exceeding 20~MeV with respect to the longitudinal position and momentum is shown in Fig.~\ref{channels}(g-i), indicating that electron acceleration occurs along the length of the laser pulse at all densities. While a bubble structure can be observed at the leading edge of the laser pulse in Fig.~\ref{channels}(d,e), the sustained strength of the ponderomotive force prevents the formation of a plasma wave or wakefield, and the electron channel density becomes almost completely depleted along the laser axis. Further, as is evident in Fig.~\ref{channels}(g-i), the majority of electrons are being accelerated within and along the cavitated channel rather then at the leading bubble structure.

\subsection{Electron motion and acceleration mechanisms}

\begin{figure}
\centering
    	\includegraphics[width=1\columnwidth]{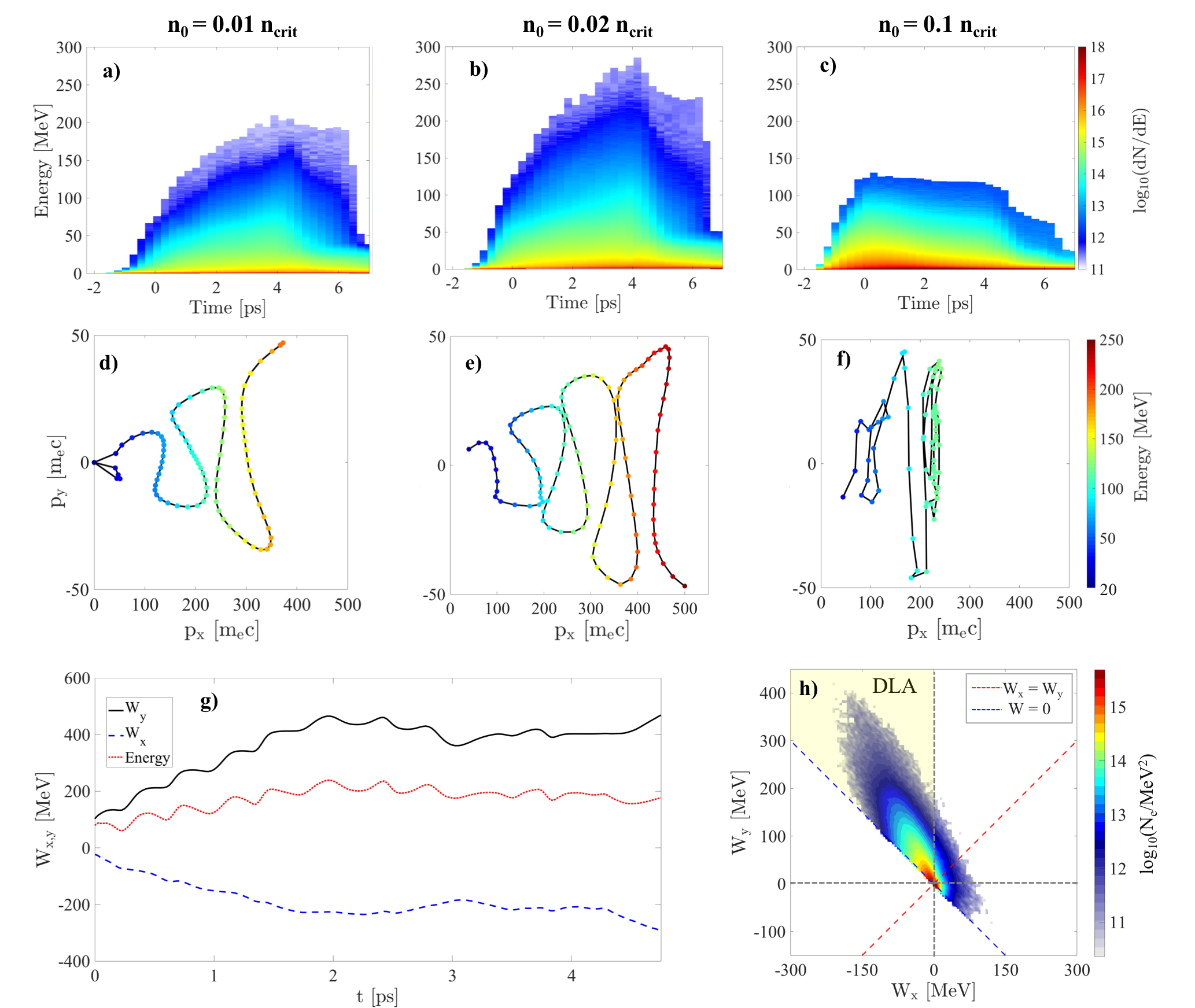}
\caption{Electron energy distribution and particle tracking from 2D PIC simulations. \textbf{a-c)} The energy distribution function of all electrons as a function of time. \textbf{d-f) }Electron trajectories in momentum space ($p_x$,~$p_y$) for randomly selected high energy electrons, from $t$ =(-0.25 - 2.75) ps, with outputs every 25 fs. \textbf{g)} Work done by the transverse and longitudinal electric fields on an electron achieving maximum energy ($E > 180$ MeV) at $t$ = 4.5 ps and $n_0$ = 0.02$~n_{crit}$.
\textbf{h) }The components of energy gain in ($W_x$,$W_y$) space at t = 4 ps for the optimal density (0.02$~n_{crit}$). The red dashed line divides the space into two regions: DLA-dominated region in the upper left and accelerated by longitudinal fields (associated with plasma waves) in the lower right.}
\label{tracking}
\end{figure}

The temporal and spatial dynamics of individual electrons provide further details on the acceleration process. The electron energy distribution in Fig.~\ref{tracking}(a-c), sampled at time intervals of 250 fs from (-2.25 to 7.25) ps, demonstrates electron energy gain from 20 MeV up to $>$ 200 MeV over 2 ps. For all densities, the electron energy saturates and the energetic electron beam exits the box at approximately 6 ps. 

To investigate the behavior of energetic electrons throughout this process, individual electron tracking was performed. Electrons with energy greater than 20 MeV, and a maximum energy 
$E > [160, 220, 110]$ MeV for $n_0$ = [0.01, 0.02, 0.1]$~n_{crit}$, respectively, at $t$ = 2.75 ps were tracked from (-0.25 - 2.75) ps, with outputs every 25 fs, to investigate differences in their trajectories close to the maximum acceleration. The momentum gain of examples of these electrons is shown in Fig.~\ref{tracking}(d-e).  For $n_0 =0.02~n_{crit}$, the electron undergoes clear periodic oscillations under the action of the laser and quasi-static channel fields, gaining energy with each cycle. At lower density, the electron is subject to weaker quasi-static channel fields, and undergoes fewer oscillations, here achieving a lower electron energy over the same period of time. At high density, the trajectory of energetic electrons is chaotic and unstable, indicating likely energy gain by stochastic processes \cite{Mangles2005} associated with self-focusing and growth of the filamentation instability. 

From the position, momentum and fields sampled by individual electrons at each time step of the simulation, the relative contributions to the total energy gain of each electron due to the transverse electric field ($E_y$) and the longitudinal electric field ($E_x$) can be calculated. The work done by $E_x$ is given by $W_x = -|e| \int_0^t E_x \cdot v_x dt'$, and is associated with plasma waves, while the work by the transverse field, $W_y$ is given by $W_y= -|e| \int_0^t E_y \cdot v_y dt'$, and is characteristic of DLA \cite{shaw2017role,ma2018ultrahigh}. At the optimal density, the temporal evolution in energy gain for an electron with energy $> 180$ MeV at $t$ = 4.5 ps is found to be dominated by $W_y$ (Fig.~\ref{tracking}g). 

To demonstrate the dominant contribution to energy gain over the entire population of electrons achieving energy $> 10$ MeV, the electron distribution in energy gain space, $(W_x, W_y)$, is plotted in Fig.~\ref{tracking}h) at $t$ =4 ps for $n_0=0.02~n_{crit}$. The majority of electrons populate the region where $W_y > W_x$, confirming DLA as the dominant acceleration mechanism, consistent with the oscillatory behavior of high energy electrons in Fig.~\ref{tracking}e). Additionally, the considerable acceleration and deceleration of electrons indicates that this process could be an efficient X-ray source. Indeed, previous work has suggested that DLA produces higher-amplitude betatron oscillations than achieved in the wakefield regime, enabling X-ray sources with much higher energies \cite{Kneip2008PRL,chen2013bright}.

\subsection{Electron dynamics under the action of channel fields}

\begin{figure}
\centering
    	\includegraphics[width=1\columnwidth]{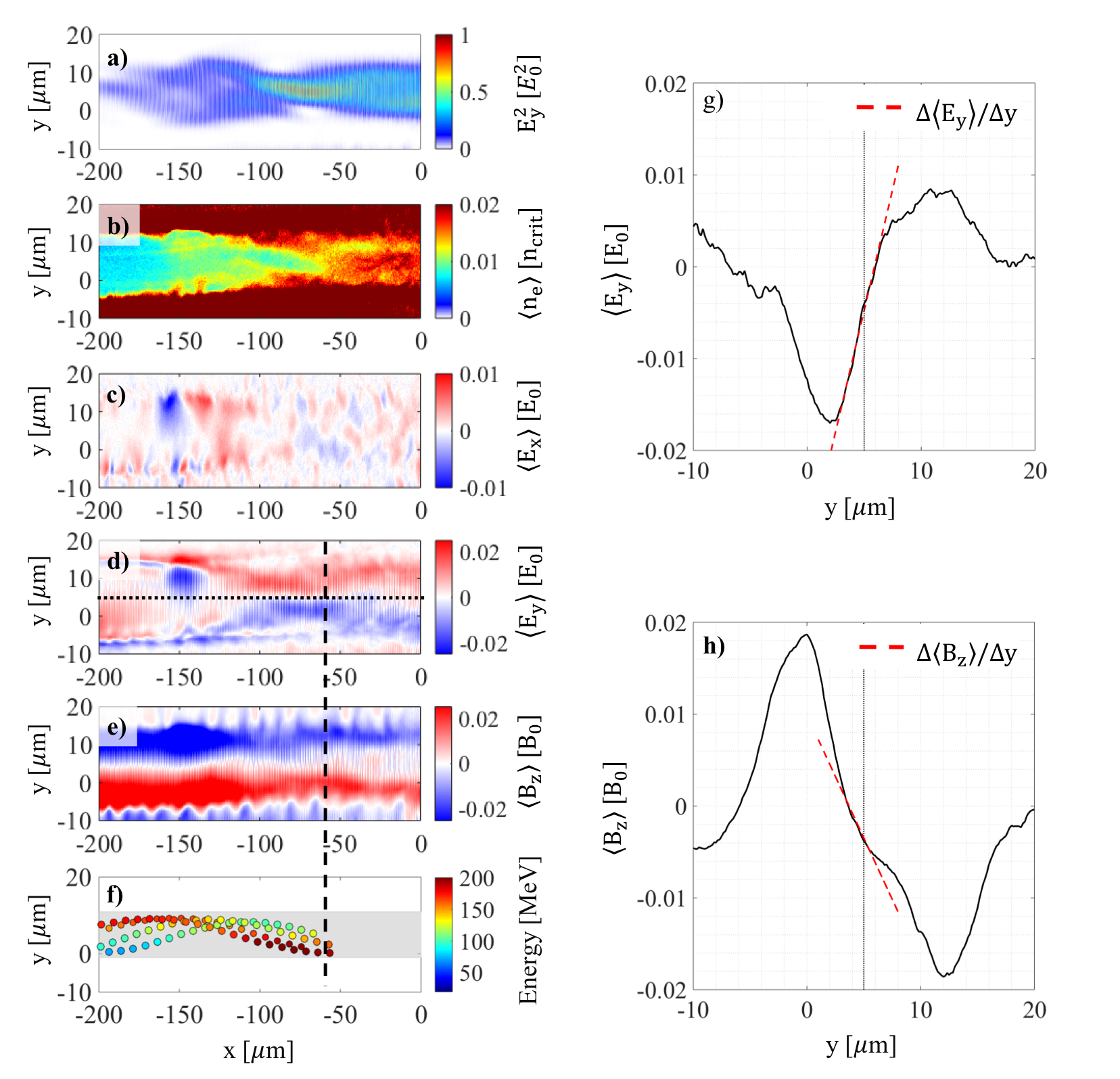}
\caption{PIC simulation snapshots of channel fields for $n_0 = 0.02~n_{crit}$ at $t = 1.75$ ps. \textbf{a)} The normalized laser intensity, with $E_y$ normalized to $E_0$. {The laser pulse extends beyond the window, spanning approximately $x=-250~\mu$m to $x=150~\mu$m. } \textbf{b)} Electron plasma density. \textbf{c)} The longitudinal electric field, $\left< E_x \right>$, normalized to $E_0$.  \textbf{d)} The transverse electric field, $\left< E_y \right>$, normalized to $E_0$. \textbf{e)} The out-of-plane magnetic field, $  \left< B_z  \right> $, representing the quasi-static channel field $B_{chan}$, normalized to $B_0$.  \textbf{f)} Location and time-history of selected high energy electrons at $t = 1.75$~ps, where the shaded region has a width of 12 $\mu$m.  \textbf{g,h)} Line-outs of $\left< E_y \right>$ and $\left< B_z  \right>$, respectively, at $x = -60~\mu$m, denoted by the vertical dashed line in (d-f). The slope of these fields is estimated by a linear fit at the center of the channel ($y = 5~\mu$m), denoted by the dotted line in d), yielding $\Delta \left< E_y \right>/\Delta y = 2.9 \times$ 10$^{10}$ statV/cm$^2$, and $\Delta \left< B_z  \right>/\Delta y = -1.5 \times$ 10$^{10}$ G/cm. Note that these snapshots have been cropped from the full simulation window to investigate the region of the pulse and channel where the highest energy electrons are located. The brackets $\left< \cdot \right>$ denote time-averaging over five laser periods.}
\label{fields}
\end{figure}

Examination of the relative strengths of the quasi-static channel fields gives insight into their effect on electron confinement and acceleration. In Figs.~\ref{fields}(a-e), snapshots of the laser field, plasma density, and time-averaged electric and magnetic fields are shown for $n_0 = 0.02~n_{crit}$ at $t = 1.75$~ps. All fields are normalized to the vacuum maximum amplitude of the laser field, denoted $(E,B)_{0}$, where $E_0$ = 5.63 $\times 10^{8}$ statV/cm and $B_0$ = 5.65 $\times 10^{8}$ G. The location and time-history of 4 tracked electrons at $t$ = 1.75 ps are shown in Fig.~\ref{fields}(f), where the shaded region has a width of 12 $\mu$m, providing a reference for the amplitude of transverse electron oscillations. These electrons, which are representative of similar high-energy electrons investigated during particle tracking, undergo clear oscillations within a confined boundary.

The fields in Figs.~\ref{fields}(c-e) are time-averaged over five laser cycles and represent the quasi-static channel fields (the averaged values are denoted using angular brackets). By visual inspection, it is clear that the longitudinal channel field, $\left< E_x \right>$, in Fig.~\ref{fields}c) is significantly weaker than the transverse field (Fig.~\ref{fields}d). This is expected during DLA, since the ponderomotive force prevents a plasma wave from forming. However, fields coinciding with density perturbations from x = [-100, 0]~$\mu$m in Fig.~\ref{fields}b) may be indicative of electron injection into the plasma channel by surface wave structures, and likely play a role in electron injection during DLA \cite{naseri2012channeling,naseri2013electron,willingale2013surface}.

The quasi-static transverse electric field, $\left< E_y \right>$, and magnetic field, $\left< B_z \right>$, result from a collective plasma response to the laser pulse. Transverse electron expulsion leads to charge separation that we characterize using a charge density $\rho_0$. The corresponding electric field (see Fig.~\ref{fields}d) reaches a maximum value of $|\left<E_y \right>|/ E_0 \approx 0.017$ at $x = -60~\mu$m. The laser pulse also drives a longitudinal electron current by pushing the plasma electrons in the forward direction. We use a current density $j_0 < 0$ to characterize this current that is typically distributed over the cross-section of the laser beam. The magnetic field sustained by the electron current is shown in Fig.~\ref{fields}e). At $x = -60~\mu$m, its maximum relative magnitude, $\left< B_z \right> / B_0 \approx 0.019$, is comparable to $|\left< E_y \right>| / E_0$.

To understand the impact of the quasi-static channel fields on the dynamics of a laser-accelerated electron, we use a standard test-particle approach where a single electron is considered in a superposition of prescribed laser and channel fields. The aim of these calculations is to provide insight into the role of the electric and magnetic fields using a simple model. The problem then reduces to solving the following equations of motion,
\begin{eqnarray}
   && \frac{d\mathbf{p}}{dt} = -|e|\mathbf{E} - \frac{e}{\gamma m_e c} [\mathbf{p} \times \mathbf{B} ],
   \label{dpdt} \\
   && \frac{d\mathbf{x}}{dt} = \frac{c}{\gamma}\frac{\mathbf{p}}{m_e c},   \label{dxdt}
   \label{dgammadt}
\end{eqnarray}
where $\mathbf{E}$ and $\mathbf{B}$ are the electric and magnetic fields acting on the considered electron, $\mathbf{x}$ and $\mathbf{p}$ are the electron position and momentum, $t$ is the time, and $\gamma = \sqrt{1 + p^2/({m_e^2 c^2})}$ is the relativistic $\gamma$-factor. This simplified model can be reasonably applied when the transverse displacement of electrons is less than the transverse size of the laser pulse, as shown in Fig.~\ref{fields}(a,f).

We approximate the laser pulse by a plane electromagnetic wave propagating along the $x$-axis with a superluminal phase velocity, $v_{ph} > c$. The superluminosity accounts for the presence of the plasma and the finite size of the channel that effectively acts as a wave-guide. The plane-wave approximation neglects the longitudinal laser electric field. This field is smaller than the transverse component roughly by a factor of $\lambda_0/R$, where $R$ is the channel radius and $\lambda_0$ is the laser wavelength in vacuum. For simplicity, we neglect the temporal change of the laser amplitude and the laser deflection observed in simulations. Then, the linearly polarized laser electric and magnetic fields can be written as \cite{wang2020electron}:
\begin{equation}
    \mathbf{E}_{wave} = E_0 \cos(\xi)~\hat{y},
\end{equation}
\begin{equation}
    \mathbf{B}_{wave} = B_0 \cos(\xi)~\hat{z},
\end{equation}
where $\xi = \omega_0(t - x/v_{ph})$ is the phase variable, $\omega_0 = 2 \pi c/\lambda_0$ is the laser frequency, and $B_0 = ({c}/{v_{ph}})E_0$. 

In order to  find the quasi-static electric and magnetic fields of the channel, we assume that $\rho_0$ and $j_0$ are constant in the channel cross-section. We also neglect their variation along $x$. We then readily find from Maxwell's equations that:
\begin{eqnarray}
     && E^{y}_{chan} = 4 \pi \rho_0 (y - y_0),
    \label{Ech} \\
    && B^z_{chan} = 4 \pi j_0 (y - y_0)/c,
    \label{Bch}
\end{eqnarray}
where the axis of the channel is located at $y = y_0$. Therefore, the total electric and magnetic fields acting on the considered electron are:
\begin{eqnarray}
    && \mathbf{E} = \mathbf{E}_{wave} + \mathbf{E}_{chan} = \left[ 0, E_{wave} + E_{chan}, 0 \right], \\
    && \mathbf{B} = \mathbf{B}_{wave} + \mathbf{B}_{chan} = \left[ 0, 0, B_{wave} + B_{chan} \right].
\end{eqnarray}

It can be verified from the equations of motion that the following quantity is conserved as the electron moves along the channel under the action of these fields:
\begin{equation}
     \gamma - \frac{up_x}{m_e c} + \frac{(y-y_0)^2}{\lambda_0^2} \left( u \kappa_B + \kappa_E \right) = \rm{C},
     \label{com-v2}
\end{equation}
where $C$ is a constant, $u \equiv v_{ph}/c$ is a normalized phase velocity, and $\kappa_{B}$ and $\kappa_E$ are two dimensionless parameters defined in terms of $j_0$ and $\rho_0$ as,
\begin{eqnarray}
    && \kappa_E \equiv ~~\frac{2 \pi |e| \rho_0 \lambda_0^2}{m_e c^2}, \\
    && \kappa_B \equiv - \frac{2 \pi |e| j_0 \lambda_0^2}{m_e c^3}.
\end{eqnarray}
The obtained conservation law is helpful in determining the amplitude of transverse electron displacements.

Typically, electrons are injected into the laser pulse from the channel walls before being accelerated. These electrons will reach the axis of the channel with an appreciable transverse momentum. It is thus appropriate to consider an electron with the following initial momentum on the axis of the channel: $p_x = 0$ and $p_y = p_i$. The constant of motion for this electron is its initial $\gamma$-factor, such that $\rm{C} = \gamma_i$. Since the longitudinal momentum and the $\gamma$-factor increase subject to the condition that $\gamma - p_x/(m_e c) > 0$, the maximum transverse displacement is achieved in the limit $\gamma - p_x/(m_e c) \rightarrow 0$. It follows from Eq.~(\ref{com-v2}) that:
\begin{equation}
    |y-y_0|_{max} = \lambda_0 \left[ \frac{\gamma_i + (u-1)\gamma}{u \kappa_B + \kappa_E} \right]^{1/2}.
    \label{r_total}
\end{equation}

In order to find $\kappa_B$ and $\kappa_E$ from our simulations, we note that:
\begin{eqnarray}
    && \kappa_E = ~~\frac{|e| \lambda_0^2 }{2 m_e c^2} \frac{\partial E_{chan}^y}{\partial y},
    \label{kE} \\
    && \kappa_B = - \frac{|e| \lambda_0^2 }{2 m_e c^2} \frac{\partial B_{chan}^z}{\partial y}.
    \label{kB}
\end{eqnarray}
In the 2D PIC simulations, $E^y_{chan}$ and $B^z_{chan}$ are represented by  $\left< E_y \right>$ and $\left< B_z \right>$, respectively. Therefore, for each field, the rate of change in $y$ can be approximated by $\Delta(\left< E_y, B_z \right>)/\Delta y$. Line-outs from $\left< E_y \right>$ and $\left< B_z \right>$ at $x=-60~\mu$m and $t = 1.75$ ps are shown in Figs.~\ref{fields}(g,h), from which the slope of these fields is estimated to be linear near the channel axis at $y \simeq 5~\mu$m (denoted by the dotted line in Figs.~\ref{fields}(d,g,h)), yielding: $\Delta \left< E_y \right>/\Delta y = 2.9 \times$ 10$^{10}$ statV/cm$^2$, and $\Delta \left< B_z \right>/\Delta y = -1.5 \times$ 10$^{10}$ G/cm. Using these values in Eqs.~(\ref{kE}) and (\ref{kB}), we obtain $\kappa_E = 0.088$ and $\kappa_B = 0.045$, indicating that $E_{chan}^y$ and $B_{chan}^z$ both play a role in transverse electron dynamics. Given that these values are comparable, it is clear that the impact of the channel magnetic field on electron dynamics is important in the considered regime of the direct laser acceleration. 

The smallest amplitude of the transverse oscillations in the considered quasi-static electric and magnetic fields is found by setting $u = 1$ and $\gamma_i = 1$ in Eq.~(\ref{r_total}), which yields:
\begin{equation}
    |y-y_0|_* = \frac{\lambda_0}{\sqrt{\kappa_B + \kappa_E}} = 2.7~\mu\rm{m}
\end{equation}
for the obtained values of $\kappa_E = 0.088$ and $\kappa_B = 0.045$. The amplitude of the transverse electron oscillations in the simulations, denoted by the shaded area behind electron trajectories in Fig.~\ref{fields}f), is noticeably wider, with a width of 12~$\mu$m ($|y-y_0|=6~\mu$m). The discrepancy can be partially attributed to a higher initial $\gamma$-factor using Equation \ref{r_total}, as it is likely comparable to $a_0 = 5$ due to the transverse injection, assuming $u=1$.

%*******************************************************************
\section{Conclusions}

Experiments and 2D PIC simulations demonstrate an optimal electron density for DLA, resulting in measurements of electron beams with energies up to (505 $\pm 75$)~MeV and up to (140 $\pm$ 30)~nC of charge. Good agreement between experimental trends and fully self-consistent 2D PIC simulations enabled investigation and diagnosis of the underlying mechanisms of DLA. The channel magnetic field was found to play an important role in defining the transverse extent of the energetic electrons, forming a boundary for electron motion with the transverse electric channel field. These observations are supported by theoretical work highlighting the profound role of a quasi-static azimuthal magnetic field on electron energy gain \textit{via} DLA \cite{pukhov1999particle,arefiev2020Bfield,gong2020forward,wang2020electron}, where much of previous work has primarily focused on channel electric fields. This result is particularly compelling for electron acceleration using longer pulse duration and higher laser intensities because magnetic fields are robust to ion motion, while electric channel fields have been shown to undergo field reversal following ion acceleration \cite{KarNJP2007}. 

This demonstration of high energy, high charge electron beams using picosecond petawatt-class laser systems could enable new applications such as positron production through the interaction of energetic electrons with a high-intensity laser pulse \cite{vranic2018extremely}, or experimental verification of the two-photon Breit-Wheeler process \cite{Pike}. Moreover, investigations into the motion of energetic electrons suggest that DLA can be used to drive bright X-ray sources with ultrashort duration \cite{Kneip2008PRL} and the capability to be accurately synchronized to short pulse laser-initiated events. Such sources could be used to image and diagnose high-energy-density physics experiments \cite{albert2016applications,wood2018ultrafast}.

\section*{Acknowledgements}
This work was supported by the National Laser Users'
Facility under grant number DE-NA0002723. Simulations for this work were performed using the EPOCH code 
(developed under UK EPSRC grants EP/G054940/1, EP/G055165/1 and EP/G056803/1) and HPC resources provided by the Texas Advanced Computing Center at The University of Texas. A.E. Hussein acknowledges funding from the National Science and Engineering Research Council of Canada and the University of California President's Postdoctoral Fellowship program. H. Chen and G. J. Williams were supported under the auspices of the U.S. Department of Energy by Lawrence Livermore National Laboratory under Contract DE-AC52-07NA27344. {The authors thank Jens Von Der Linden at Lawrence Livermore National Laboratory for assistance with the calibration of the electron positron spectrometer.}

\section*{References}
\bibliographystyle{unsrt}

\end{document}